\documentclass[sigconf]{acmart}

\usepackage{color}
\usepackage{algorithm}
\usepackage{algorithmic}
\usepackage{graphicx}
\usepackage{amsmath}
\usepackage{multirow}
\usepackage{balance} 

\AtBeginDocument{%
  \providecommand\BibTeX{{%
    \normalfont B\kern-0.5em{\scshape i\kern-0.25em b}\kern-0.8em\TeX}}}

\copyrightyear{2021}
\acmYear{2021}
\setcopyright{acmcopyright}
\acmConference[CIKM '21]{Proceedings of the 30th ACM
International Conference on Information and Knowledge Management}{November
1--5, 2021}{Virtual Event, QLD, Australia}
\acmBooktitle{Proceedings of the 30th ACM International Conference on Information
and Knowledge Management (CIKM '21), November 1--5, 2021, Virtual Event, QLD,
Australia}
\acmPrice{15.00}
\acmDOI{10.1145/3459637.3482292}
\acmISBN{978-1-4503-8446-9/21/11}

\begin{document}

\title{Reinforcement Learning to Optimize Lifetime Value in Cold-Start Recommendation}

\author{Luo Ji}
\affiliation{%
  \institution{DAMO Academy, Alibaba Group}
  \city{Hangzhou}
  \country{China}
}
\email{jiluo.lj@alibaba-inc.com}

\author{Qi Qin}\authornote{The first two authors contributed equally to this research.}
\authornote{This work was done when Qin Qi was an intern at Alibaba.}
\affiliation{%
  \institution{Center for Data Science, AAIS, Peking University}
  \city{Beijing}
  \country{China}}
\email{qinqi@pku.edu.cn}

\author{Bingqing Han}
\affiliation{%
  \institution{DAMO Academy, Alibaba Group}
  \city{Hangzhou}
  \country{China}
}
\email{bingqing.hbq@alibaba-inc.com}

\author{Hongxia Yang}\authornote{Corresponding author}
\affiliation{%
  \institution{DAMO Academy, Alibaba Group}
  \city{Hangzhou}
  \country{China}
}
\email{yang.yhx@alibaba-inc.com}

\renewcommand{\shortauthors}{Ji and Qin, et al.}

\begin{abstract}
Recommender system plays a crucial role in modern E-commerce platform. Due to the lack of historical interactions between users and items, cold-start recommendation is a challenging problem. In order to alleviate the cold-start issue, most existing methods introduce content and contextual information as the auxiliary information. Nevertheless, these methods assume the recommended items behave steadily over time, while in a typical E-commerce scenario, items generally have very different performances throughout their life period. In such a situation, it would be beneficial to consider the long-term return from the item perspective, which is usually ignored in conventional methods. Reinforcement learning (RL) naturally fits such a long-term optimization problem, in which the recommender could identify high potential items, proactively allocate more user impressions to boost their growth, therefore improve the multi-period cumulative gains. Inspired by this idea, we model the process as a Partially Observable and Controllable Markov Decision Process (POC-MDP), and propose an actor-critic RL framework (RL-LTV) to incorporate the item lifetime values (LTV) into the recommendation. In RL-LTV, the critic studies historical trajectories of items and predict the future LTV of fresh item, while the actor suggests a score-based policy which maximizes the future LTV expectation. Scores suggested by the actor are then combined with classical ranking scores in a dual-rank framework, therefore the recommendation is balanced with the LTV consideration. Our method outperforms the strong live baseline with a relative improvement of $8.67\%$ and $18.03\%$ on IPV and GMV of cold-start items, on one of the largest E-commerce platform.
\end{abstract}




\maketitle

\section{Introduction}

\begin{figure}[t]
  \centering
  \includegraphics[width=8cm]{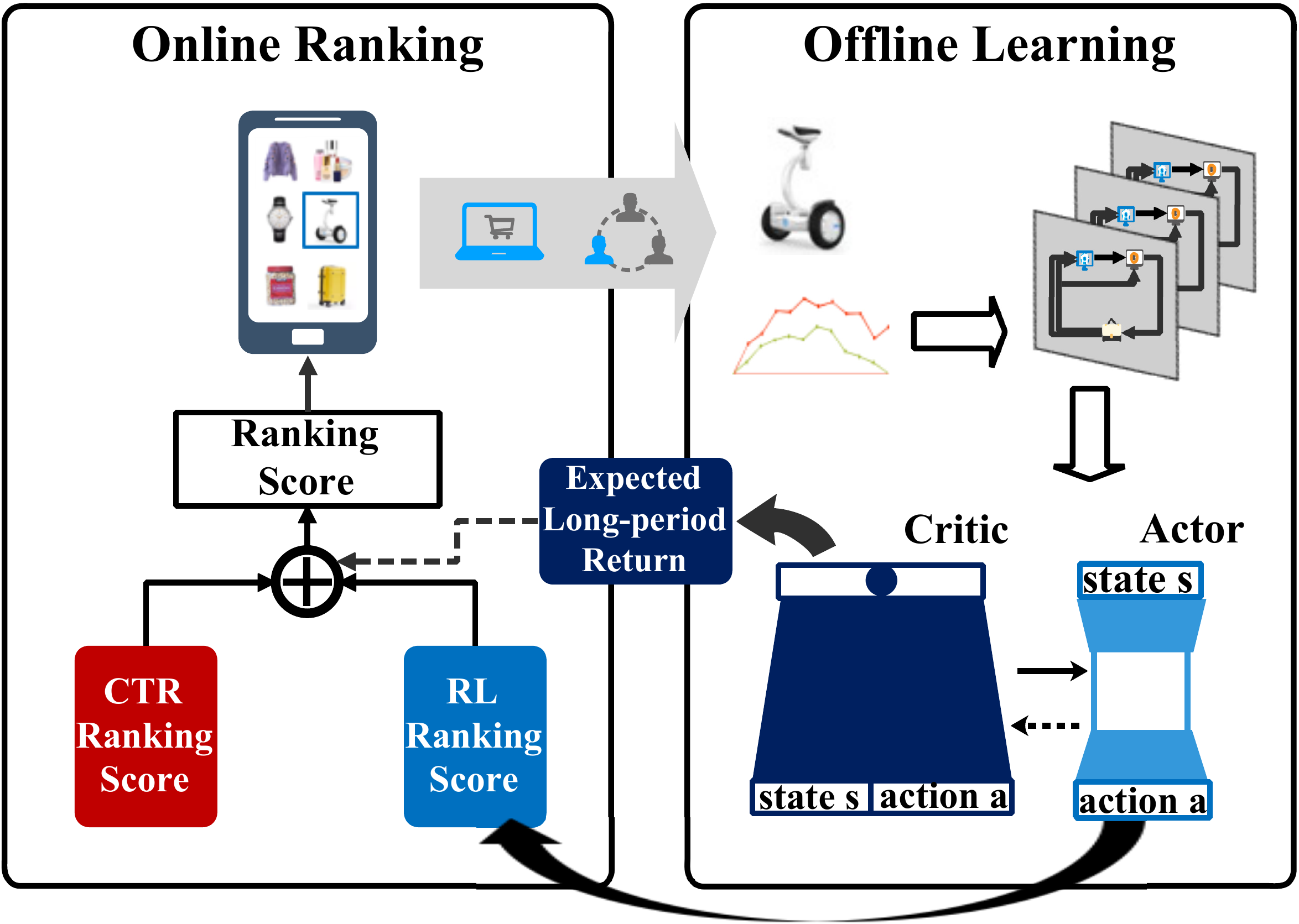}
  \caption{Illustration of the RL-LTV algorithm framework. An RL is employed to solve a item-level MDP and provides a policy optimizing item long-term rewards (LTV). The policy score is combined with CTR ranking score with their combination weight adjusted by the critic expectation. The ultimate ranking score is applied in online ranking.}
  \label{global_framework}
\end{figure}


Recommender systems (RS) have become increasingly popular and have been utilized in a variety of domains (\emph{e.g.} products, music, movies and etc.) \cite{Anidorif2015Recommender}. RS assists users in their information seeking tasks by suggesting a list of items (\emph{i.e.}, products) that best fits the target user's preference. For a practical system, a common pipeline is that a series of items are first retrieved from enormous candidates, then sorted by the ranking strategy to optimize some expected metric such as CTR (\emph{i.e.,} Click-Through Rate) \cite{2011Unbiased}. 

However, due to the lack of user-item interactions, a common challenge is the cold-start recommendation problem \cite{2019Addressing}. Solution to the cold-start problem may depend on the platform characteristics. Traditional way to solve the cold-start problem is leveraging auxiliary information into the recommendation systems (\emph{e.g.}, content based \cite{2016Latent, 2016Collaborative}, heterogeneous information \cite{Shi2016A,2020Meta} and cross-domain \cite{2018Li, 2020CDLFM}). Although they have achieved good performance, they focus on the instant reward, while the long-term rewards is ignored. Correspondingly, there is recently increasing attention on the long-term/delayed metric, and solution to optimize the long-term user engagement \cite{wu2017returning, 2019UserEngage} is proposed. However, a long-term viewpoint from the item aspect is still missing. 

In the case of E-commerce, where the recommended items are typically products, there is a clear need to consider their long-term behaviors, which are changing throughout their life periods. The life dynamics of products share a similar development pattern, as stated in \textit{product life theory} \cite{Levitt1965PLC, Hui2012PLC}. \cite{he2018speeding} further proposes four distinct stages including \textit{introduction}, \textit{growth}, \textit{maturity} and \textit{decline}, and uses a mathematical tool to model the product life stages and predict the transit probability between different stages. In this paper, we also consider the product life dynamics, in a continuous, numerical manner. Within the scope of cold-start recommendation, items on the earlier stages (\emph{i.e.}, \textit{introduction} and \textit{growth}) are paid more attention in this work. For a recently introduced item, recommendation algorithms focus on the instant metric may have the exposure bias. Typically, such fresh items are probably not preferred by the instant metric due to the lack of historical behavior, therefore they are subject to low ranking preference. As a result, there might be severe Matthew Effect in the existence of conventional algorithms, in which the mature items keep receiving more impressions and the fresh items are hard to grow up. However, for some fresh items, they could be more and more popular with some investment of user impressions, and yield more returns in the future. From this aspect, a smart cold-start recommender should be able to identify high potential items in advance, and assign them higher ranking priority; while a low potential product could be penalized in contrary. A multi-period maximization task is then needed.

Reinforcement Learning (RL) provides a natural, unified framework to maximize the instant and long-term rewards jointly and simultaneously. Nevertheless, considering the complexity of actual online environment, building an interactive recommender agent between user and item, as well as considering the long-term rewards is a challenging task. The trial-and-error behavior of RL might harm the system performance, or affect the user satisfaction. The calculation complexity could also be prohibitively expensive. From these considerations, and considering the fact that the time-evolution of products are naturally in a much slower time scale than online recommendation (days VS milliseconds), in this study we instead define an off-policy learning method on the item level and on the daily basis, which makes the solution practical.

In this paper, we proposes a novel methodology, named \textit{reinforcement learning with lifetime value} (RL-LTV), to consider the long-term rewards of recommended items inside the cold-start recommendation problem. Such long-term rewards are called by item \textit{lifetime values} (LTV) in this paper. An off-policy, actor-critic RL with a recurrent component is employed to learn the item-level dynamics and make proactive long-term objected decisions. Information of aforementioned product life stages are encoding by the recurrent hidden memory states, which are studied by a LSTM ~\cite{hochreiter1997long} component, shared by actor and critic. To transfer the information from historical items to cold-start items, we introduce item inherent features, trending bias term, and memory states as extra inputs into both the actor and critic. One of the most important action output by the actor, the item-infinity score of LTV, is then incorporated with the conventional ranking score to form a dual-rank framework. Their combination weight is determined by action-value suggested by the critic. Figure \ref{global_framework} illustrates the entire framework of our proposed algorithm. 

The major contributions of this paper are as follows:
\begin{itemize}
\item We define a concept of \textit{Partially Observable and Controllable Markov decision process} (POC-MDP) to formulate the product life dynamics. Unobservable states depict the intrinsic life stages, while uncontrollable states can affect product growth speed but are independent of actions.

\item We incorporate the item LTVs into the online ranking by RL-LTV. By prioritizing high potential cold-start items during the ranking, the exposure bias of cold-start items could be overcome. To the best of our knowledge, this is the first time that such a technique is applied to solve the cold-start recommendation problem.

\item Knowledge of mature items could be generalized and transferred to cold-start items, even for those first-introduced items. To achieve this, we build the MDP on the item-level, with continual observation and action spaces, as well as parameter-shared policy and critic networks.

\item We design a learning framework called IE-RDPG to solve a large-scale RL problem in an itemwise, episodic way. The algorithm is deployed into production and improvements of $8.67\%$ and $18.03\%$ on IPV and GMV for cold-start items are observed.
\end{itemize}

The rest of the paper is organized as follows. The connection with previous works is first discussed in Section \ref{sec:related_work}. Preliminaries are then introduced in Section \ref{sec:model_def}. The POC-MDP formulation and its learning algorithm are stated in Section \ref{sec:model_framework}. Experiment results are summarized in Section \ref{sec:experiment}. Finally Section \ref{sec:conclusion} concludes this paper.

\section{RELATED WORK}
\label{sec:related_work}

In this section, we will briefly review representative works of cold-start recommendation and reinforcement learning. 

\textbf{Cold-start Recommendation:} Although collaborative filtering and deep learning based model has achieved considerable success in recommendation systems\cite{bokde2015role,2017Collaborative}, it is often difficult to deal with new users or items with few user-item interactions, which is called cold-start recommendation. The traditional solution of cold-start recommendation is to introduce auxiliary information into the recommendation system, such as content-based, heterogeneous information and cross domain. Specifically, the content-based methods rely on data augmentation by merging the user or item side information \cite{2016Latent,2016Collaborative,Zhang_2019, 2019Addressing}. For example, \cite{2016Latent} presents an approach named visual-CLiMF to learn representative latent factors for cold-start videos, where emotional aspects of items are incorporated into the latent factor representations of video contents. \cite{2016Collaborative} proposes a hybrid model in which item features are learned from the descriptions of items via a stacked diagnosing auto-encoder and further combined into a collaborative filtering model to address the item cold-start problem. In addition to these content-based features and user-item interactions, richer heterogeneous data is utilized in the form of heterogeneous information network \cite{2019Metagraph,2011PathSim,2020Meta} , which can capture the interactions between items and other objects. For heterogeneous information based methods, one of the main tasks is to explore the heterogeneous semantics in recommendation settings by using high-order graph structure, such as Metapath or Metagraph. Finally, cross-domain methods based on transfer learning, which applies the characteristics of the source domain to the target domain \cite{2018Li,2020CDLFM}. The premise of this type method is that the source domain is available and users or items can be aligned in the two domains. For example, \cite{2018Li} presents an innovative model of cross-domain recommendation according to the partial least squares regression (PLSR) analysis, which can be utilized for better prediction of cold-start user ratings. \cite{2020CDLFM} propose a cross-domain latent feature mapping model, where the neighborhood-based cross-domain latent feature mapping method is applied to learn a feature mapping function for each cold-start user. 

Although these methods have achieved good performance, most of them only alleviate the cold-start problem from the single-period viewpoint while ignore the long-term rewards. Recently, there is some study which tries to study the long-term effect from the user-side \cite{2019UserEngage}, but not the item side. In this paper, we try to solve a item cold-start problem by not only using items content information, but also determine the ranking preference of items according to their long-term returns.

\textbf{Reinforcement Learning:} Our approach connects to the widely application of reinforcement learning on recommendation problems. These applications include different strategies: value-based \cite{Chen2018StabilizingRL,Taghipour2018,2018DRN}, policy-based \cite{Hau1997, 2013PEGASUS,Gong2019ExactKRV,Chen2019TopKOC}, and model-based \cite{Bai2019ModelBasedRL} methods. When the environment is identified as a partial observable MDP (POMDP), the recurrent neural network is a natural solution to deal with the hidden states \cite{Bakker2001ReinforcementLW,Zhao2018RecommendationsWN,2018Learning,2019UserEngage}. Reinforcement Learning also helps to generate an end-to-end listwise solution \cite{zhao2019deep} or even jointly determines the page display arrangement \cite{Zhao_2018}.

There are also RL-based studies for cold-start recommendation. \cite{wang2020offline} proposes an offline meta level model-based method. \cite{ding2017coldstart} combines policy-gradient methods and maximum-likelihood approaches and then apply this cold-start reinforcement learning method in training sequence generation models for structured output prediction problems. \cite{2019UserEngage} uses reinforcement learning to solve the multi-period reward on user engagement. \cite{he2018speeding} proposes a RL-based framework for impression allocation, based on consideration of item life period stages. In their work, the item stages are explicitly identified and predicted by a mathematical model while RL is used inside the impression allocation problem. 

In contrast to \cite{he2018speeding}, in this paper the item stage information is implicitly studied by reinforcement learning in an end-to-end manner. In particular, we use recurrent neural network to encode the hidden state as a continual and dense representation of life stages, based on item histories. Our framework jointly studies this recurrent hidden state, the action value as a prediction of item long-term reward, as well as the ranking policy.




\section{Preliminaries}
\label{sec:model_def}

In this section, we first introduce the background of product metabolism and the idea of item lifetime value on E-commerce, which basically motivates this work. Based on the understanding of this scenario, we define a special type of Markov Decision Process to model the metabolism of an item. The basic DDPG algorithm is finally shown as the underlying learning approach.


\subsection{Item Metabolism and Lifetime Value}
\label{sec:item_growth}

\begin{figure}
  \centering
  \includegraphics[width=8cm]{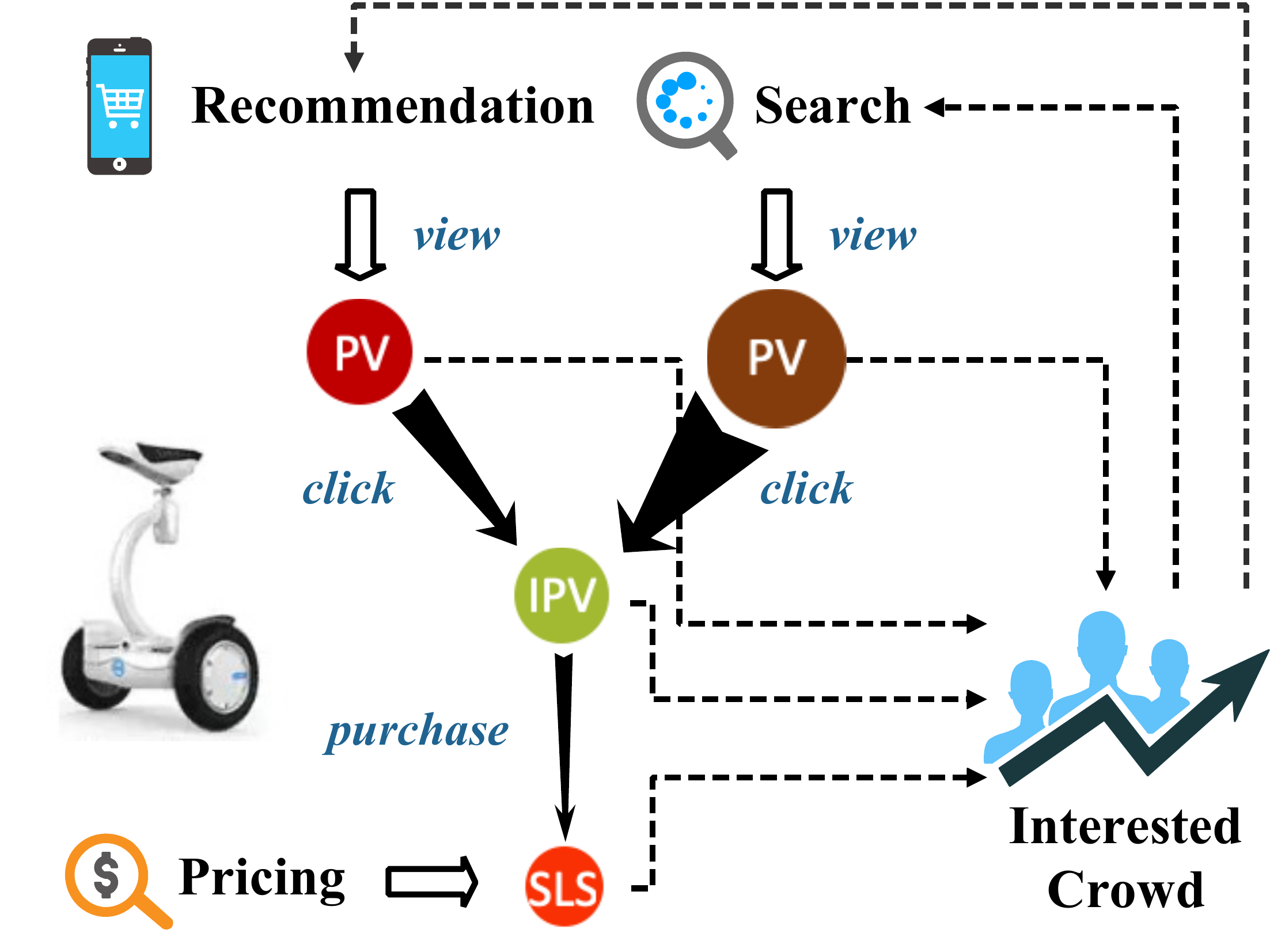} 
  \caption{Item metabolism in e-commerce. Recommendation and Search feed the page views of item and determine the click-through rates. Pricing strategy further determines how many sales are converted. All of these behaviors result in more people interested with the item and choose to search it in the future. The item then obtains more priority in search and recommendation algorithms because of more historical information.}
  \label{item_growth}
\end{figure}

A typical pattern of product metabolism on the e-commerce platform is shown in Figure \ref{item_growth}. Assuming a balancing electric scooter is just introduced into the platform, it usually receives few user interest and the statistics are in low-level. Although the item can be left to grow up by itself, several channels could be utilized to change its situation. First, the item's user exposure (or more specifically, the page views (PV)) is significantly affected by both search and recommendation algorithms. Second, reasonability of search and recommendation directly affects the click-through rate (CTR), \emph{i.e.}, how many item page views (IPV) could be clicked from PV. Furthermore, as the third channel, the pricing strategy from the pricing system (PS) has substantial impact that how many number of sales (SLS) are converted from IPV. During this process, PV, IPV and SLS are accumulated,  the item's reputation is built, and the interested group of users continues to expand. As more users appeal to the item, their behaviors help the specific item claim more importance in search and recommendation algorithms, therefore a positive feedback closed-loop mechanism could be created. As item elapses, the fresh item may have growing life dynamics trajectories of key indicators, including PV, IPV and SLS. 

However, not all items can achieve such a positive closed-loop mechanism. 
The growth rate of an item depend on its inherent characteristics (what the item is), brand, market, external trends or noise, and the algorithms. Actually, most items finally fall into the group called "long-tail products", with few user views, clicks or purchases. Therefore, it would help if one could identify if the item could return significant future PVs, IPVs and GMVs \footnote{The gross merchandise value from the item, which equals to $SLS$ times the averaged paid price.}, such that the star products and the long-tail products can be classified even at their early life stage. In this paper, we call such long-term rewards as the item's Lifetime Value (LTV). By allocating more resources for those high potential products, the platform would be repaid with more LTV in the future, and makes the entire ecosystem grow and prosper. As shown in Figure \ref{item_growth}, search, recommendation and pricing are possible important tools to allocate the resource and adjust the item dynamics.





\subsection{MDP and its Extensions}
\textit{Markov Decision Process} (MDP) is typically employed to model the sequential decision making problem. Nominal MDP usually consists of the 4-tuple $(\mathcal{S, A, R, P})$, where $\mathcal{S, A, R, P}$ are the state space, action space, set of rewards, and transition probability functions, respectively. At time step $t$, the state $s_t \in \mathcal{S}$ represents the current system stage, which is affected by an action $a_t \in \mathcal{A}$ from the agent, generating a reward $r_t$ by the reward function $\mathcal{S} \times \mathcal{A} \to \mathcal{R}$, as well as the next state $s_{t+1} \in \mathcal{P}(s_t, a_t)$. For such nominal MDP, it assumes that all elements of $s$ can both be observed by agent, and be changed by the agent's action. 

However, it is rare that this assumption holds in the real environment. For example, the Pong game in Atari have both the current image and the ball velocity as state, while only the former is provided to the agent. \textit{Partially Observable Markov Decision Process} (PO-MDP) captures the partial observability part of system complexity. It is instead described as a 5-tuple $(\mathcal{S, A, P, R, O})$, in which $\mathcal{O}$ denotes the set of observable states, \emph{i.e.}, $\mathcal{O}$ is subset of $\mathcal{S}$. 

Another form of system complexity is relatively less studied, the MDP \textit{with uncontrollable states} \cite{Arruda2009StandardDP,Liang_2019}. In such situations, some elements of $s_t$ can never be affected by agent actions, but they can determines $s_{t+1}$ therefore affect future rewards too. We here pay especial attention to such form of uncontrollablity, which will be discussed with more details in the next section. 

In our topic, it is believed that both unobservable and uncontrollable states exist, both of which are discussed with more details in Section \ref{sec:model_framework}. 
We deal with the above concerns and define a \textit{Partially Observable and Controllable Markov Decision Process} (POC-MDP), which literally means there are some unobservable states and some uncontrollable states in MDP at the same time. Although the term "state" can denote all states no matter they are observable or controllable, for clarity, we use the notation $s$ to present only the nominal (both observable and controllable) states. The unobservable states (but controllable) are denoted by $h \in \mathcal{H}$, and the uncontrollable (but observable) states are denoted by $x \in \mathcal{X}$. As a result, our POC-MDP now has the 6-tuple $(\mathcal{S, A, P, R, O, H})$. Note now the observation is the concatenation of nominal states and uncontrollable states\footnote{In this paper, we use the square bracket to represent the concatenation of multiple variables, for simplicity.}, \emph{i.e.}, $o := [s, x]$.


\subsection{The DDPG method}
\label{sec:ddpg}

In this section we give a brief introduction to an model-free, off-policy, actor-crtic RL algorithm, the Deep Deterministic Policy Gradient (DDPG) method \cite{lillicrap2019continuous}, which is closely related to our proposed approach.


In DDPG, the actor network $\pi$ is approximated by the net parameter $\theta$ and generates the policy, and the critic network $Q$ is approximated by the net parameter $w$ and generates the action-value function. Gradients of the deterministic policy is
\begin{equation}
\triangledown_{\theta} J = \mathbb{E}_{s \backsim d^{\pi}}[\triangledown_{\theta} \pi_{\theta}(s)  \triangledown_{a}Q_w(s,a)|_{a=\pi(s)}]
\notag
\end{equation}
where $d^{\pi}(s)$ is a discounted distribution of state $s$ under the policy of $\pi$. Then $\theta$ can be updated as 
\begin{equation}
\theta \leftarrow \theta + \eta \mathbb{E}_{s \backsim d^{\pi}}[\triangledown_{\theta} \pi_{\theta}(s)  \triangledown_{a}Q_w(s,a)|_{a=\pi(s)}]
\notag
\end{equation}
with $\eta$ as the learning rate. For the critic, the action-value function can be obtained iteratively as
\begin{equation}
Q_w(s_t, a_t) = \mathbb{E} [r_t + \gamma \mathbb{E}_{a \backsim \pi_{\theta}} [Q_w(s_{t+1}, a_{t+1})]]
\notag
\end{equation}
$w$ is updated by minimizing the following objective function
\begin{equation}
\small
\mathop{\text{min}}_{w} L = \mathbb{E}_{s \backsim d^{\pi}}[(R(s_t,a_t) + \gamma Q_{w^{'}}(s_{t+1},\pi_{\theta^{'}}(s_{t+1})) - Q_{w}(s_{t},\pi_{\theta}(s_{t})))^2] 
\notag
\end{equation}
where $\pi_{\theta^{'}}$ and $Q_{w^{'}}$ are the target networks of actor and critic. The target network parameters can be softly updated  by 
\begin{align}
\theta^{\mu^{'}} {\leftarrow} \tau {\theta^{\mu^{'}}} + (1-\tau)\theta^{\mu} \notag \\
w^{'} \leftarrow \tau w^{'} + (1-\tau)w \notag
\end{align}






\section{Method}
\label{sec:model_framework}

This section illustrates the key concepts of our approach, including how we implement our \textit{RL-LTV}, and how its action is applied in the online system. First definitions of terms in POC-MDP are listed, then architectures of the actor and critic networks are introduced, after which the learning algorithm follows. The actor outputs a preference score, which is linearly combined with the ranking score from a conventional CTR model in a dual rank framework. The new ranking score is then applied to sort items within the response to each request. Table \ref{tab:note} summarizes important symbols which are frequently used in the current and related sections.

\begin{table}
 \caption{Notations.}
 \label{tab:note}
 \scalebox{0.75}{
  \begin{tabular}{cl}
    \toprule
    Notation & Description\\
    \midrule
    $x_i$ & the item inherent, time-invariant features \\
    $x_t$ & the time-variant, trending-bias factors \\
    $\mathcal{X}$, $x$ & uncontrollable state space $\mathcal{X}$, $x := [x_i, x_t] \in \mathcal{X}$ \\
    $\mathcal{S}$, $s$ & state space $\mathcal{S}$, $s \in \mathcal{S}$ \\
    $\mathcal{O}$, $o$ & observation space $\mathcal{O}$, $o := [s, x] \in \mathcal{O}$ \\
    $\mathcal{H}$, $h$ & unobserable, hidden state space $\mathcal{H}$, $h \in \mathcal{H}$ \\
    $\mathcal{A}$, $a$ & action space $\mathcal{A}$, $a \in \mathcal{A}$ \\
    $\mathcal{R}$, $r$ & reward space $\mathcal{R}$, $r \in \mathcal{R}$ \\ 
    $\gamma$ & discount factor to balance immediate and future rewards \\
    $J$ & discounted cumulative return\\
    $\mathcal{T}$ & the transition $(o_t, a_t, r_t, o_{t+1})$\\
    $\pi_{\theta}(a|o)$ & policy function with $\theta$ as actor parameters\\
    $Q_{w}(o, a)$ & state-action value function with $w$ as critic parameters\\
    $PV$ & page views of all channels (search plus recommendation)\\
    $PV_{\text{rec}}$ & page views from solely recommendation\\
    $IPV$ & item page views jumped from all channels \\
    $SLS$ & number of sales \\
    $p$ & the discount degree of (averaged) paid price \\
    $GMV$ & the revenue of item, basically price times SLS \\
    $y_{ctr}$& the pointwise user-item affinity score from CTR model \\
    $y_{rl}$& the item-affinity score from RL-LTV \\
    $y$& the finally-applied weighted pointwise recommendation score \\
    $\alpha$ & linear weight between $y_{ctr}$ and $y_{rl}$\\ 
  \bottomrule
\end{tabular}
}
\vspace{-2mm}
\end{table}


\subsection{Definitions}
\label{subsec:definitons}

\begin{figure}[t]
  \centering
  \includegraphics[width=8cm]{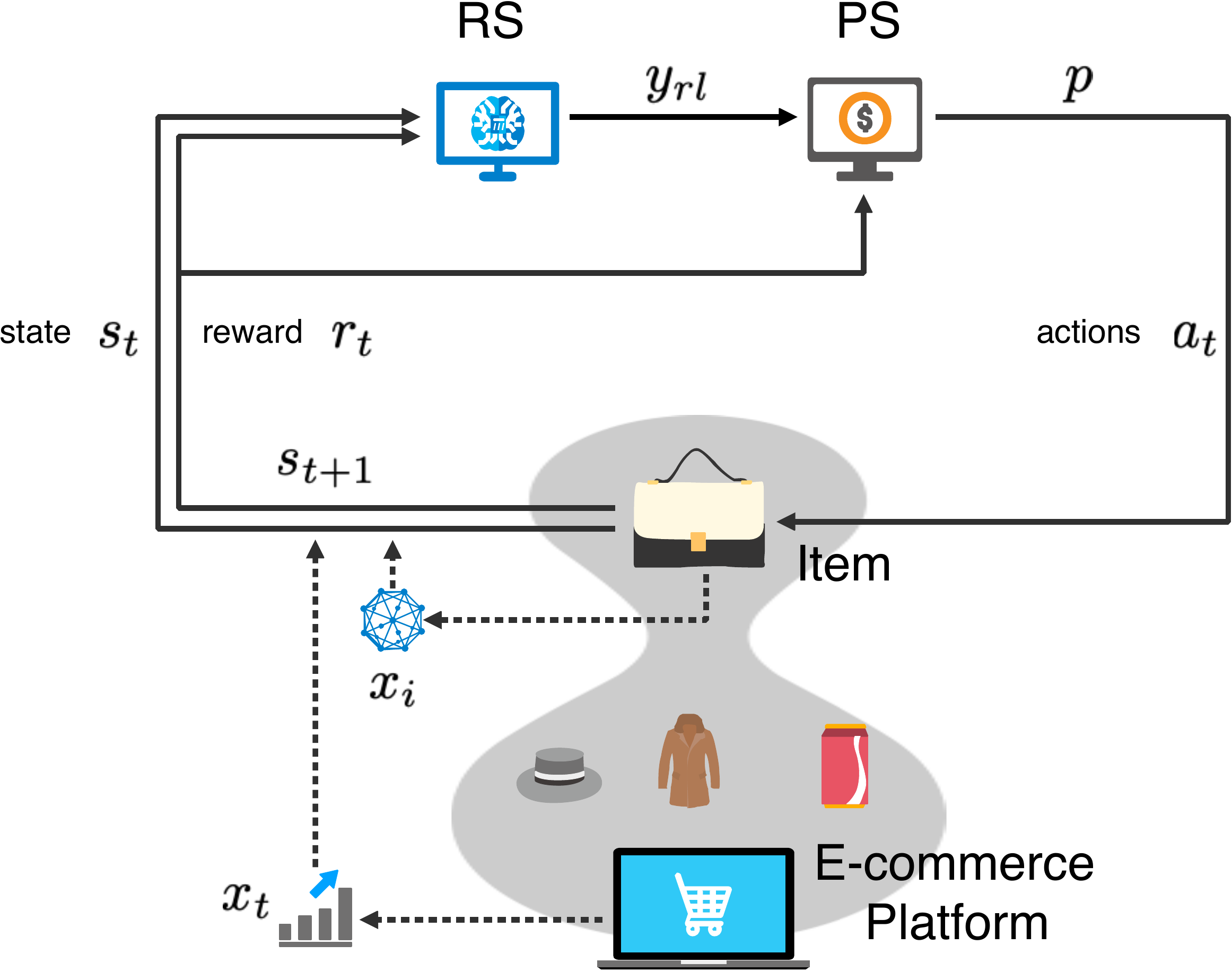}
  \caption{Schematic representation of POC-MDP. The agent are two-level in which $y_{rl}$ is the action of RS and $p$ is the action of PS. The item inherent, item-invariant feature is denoted by $x_i$, and the trending bias, time-variant factor is denoted by $x_t$, both of which can affect the dynamics.}
  \label{mdp}
\end{figure}

In this paper, we consider a two-level agent, including RS which affects the click rate and PS which affects the conversion rate. The environment is the e-commerce ecosystem. Figure \ref{mdp} provides a snapshot of our POC-MDP, with terms defined below:

\textbf{State}. The state space is defined on the item-level, representing the observable part of current item life stage, including the item's time on the market,  PV (both current and accumulated), IPV (both current and accumulated), SLS (both current and accumulated), and properties (number, averaged activeness frequency, averaged purchasing power, \emph{etc}) of the user crowd currently interested with the item. 


\textbf{Action}. Action is denoted by $a_t = [y_{rl, t}, p_t] \in \mathcal{A}$. For the RS part, $y_{rl} \in (0,1)$ indicates the RS's preference ranking score for a specific item. PS at the downstream of RS has the action of $p$ which is modelled as the price discount degree, the averaged paid price divided by the original price. Both $y_{rl}$ and $p$  can be easily retrieved from the real data.


\textbf{Uncontrollable State}. We formulate such states from the inspiration that an item's next state is not only result of its current state, but also determined by its inherent feature and extrinsic trending-bias factors. The inherent feature
(\emph{e.g.,} title, cover image, category, brand, shop name) can always be assumed item-invariant. Therefore, we denote them as $x_i$ (where i means both inherent and item-invariant). Notably, within $x_i$ a pre-trained, 128-dim item title \& image multi-modal embedding is employed as the core information to bridge the information between different items \cite{InterBERT}. Other parts of $x_i$, including category, brand and shop, are input as ID-type features. The extrinsic bias factors can have different trend sources, including the entire market, the platform campaign, item seller, brand or category. We denote them as $x_t$ because obviously they are time-variant. For each trending source, we include their moving average growth percentage (MAGP) of PV, IPV and SLS. 

\textbf{Reward}. As mentioned before, the MDP is defined on the item-level in order to share information between items. However, online recommendation contains a listwise ranking which means different item competes with each other for impression allocation. Considering this dimensional mismatch, we can not simply define the reward as a absolute value (LTV of PV, IPV, SLS, or their linear combinations) because in this case RL would simply generate $y_{rl}$ for each item as large as possible. Instead, we choose the form of reward to be similar with ROI(the return of investment), which is defined as the return of the next-step IPV with the investment of current recommended PV:

\begin{equation}
\small
   r_t = \frac{\text{IPV}_{t+1}}{\text{PV}_{rec,t}}
\end{equation}
\vspace{-1mm}


\textbf{Long-term Gain and Discount Factor.} The multi-period, cumulative gain is defined by 
\begin{equation}
\label{eq:v}
   J_t = \sum_{i=0}^{T} \gamma^{i} r_{t+i} 
\end{equation}
which could be viewed as our ROI version of LTV. $T$ is the window length and $\gamma$ is the important discount factor, both of which control how far a model looks forward when optimizing $J$. In our study, $\gamma$ is set to a smaller number, \emph{i.e.}, $0.5$, indicating that our RL balances two sides of considerations: (1) the long-term total rewards, as well as (2) to speed up the growth of cold-start items as soon as possible.





\subsection{Actor}

The actor is designed to generate the action $a_t = [y_{rl}, p]$ given the current observation $o_t = [s_t, x_t, x_i]$. Such policy function is approximated by the actor net $\pi_{\theta}(a_t|o_t)$ with $\theta$ representing net parameters.  As stated in Subsection \ref{subsec:definitons}, the item inherent feature $x_i$ including two parts: the first part is the pretrained embedding vector; the second part is composed of different IDs, all of which are feed into an encoder network $f_{e}(x_i, w_e)$. Embedding generated by this encoder is concatenated with the pretrained vector and forms $x_{e,i}$, the embedded version of $x_i$. Then we have the new observation after embedding $o_{e,t} = [s_t, x_t, x_{e,i}]$. 
Meanwhile, in order to allow the agent to capture the intrinsic item life stage, we introduce an LSTM cell to encode the sequential information of historical observations into continuous hidden states:
\begin{equation}
\begin{split}
    & \mathnormal{Z}_{f,t} = \mathnormal{\sigma} (W_f \cdot [h_{t-1}, o_{e,t}]) \\
    & \mathnormal{Z}_{u,t}  = \mathnormal{\sigma} (W_u \cdot [h_{t-1}, o_{e,t}]) \\
     & \mathnormal{Z}_{o,t} = \mathnormal{\sigma} (W_o \cdot [h_{t-1}, o_{e,t}]) \\
   & \tilde{c}_t  = \mathnormal{tanh} (W_c \cdot [h_{t-1}, o_{e,t}]) \\
   &  c_t =\mathnormal{Z}_{f,t} \cdot c_{t-1} + \mathnormal{Z}_{u,t} \cdot  \tilde{c}_t \\
    &  h_t= \mathnormal{Z}_{o,t}  \cdot  \mathnormal{tanh}(c_t). 
\end{split}
\end{equation}
where $\mathnormal{Z}_{f,t}, \mathnormal{Z}_{u,t}, \mathnormal{Z}_{o,t}$ are the forget, update, output gates. $W_f$, $W_u$,
$W_o$, $W_c$ are LSTM parameters. $c_t$ and $h_t$ are the cell state and hidden state, respectively. In general, actor's inputs are $[o_{e,t},h_t]$.   

In order to enhance the actor's ability to memory $o_{e,t}$ and generalize $h_t$, a wide and deep structure \cite{Cheng2016WideD} is applied:
\begin{equation}
    \begin{split}
        o_{wide} &= f_{wide}(W_w,o_{e,t}),\\
        o_{deep} &= f_{deep}(W_d,[o_{e,t},h_t])
    \end{split}
\end{equation}
in which $f_{wide}$ and $f_{deep}$ are one or three fully connected layers, respectively. To obtain the first action (\emph{i.e.}, $y_{rl}  \in (0,1)$), we use the following functions

\begin{equation}
\begin{split}
    & S  = \text{softmax}(W_s^T \cdot [o_{deep}, o_{wide}]);\\
     & y_{rl} = \sigma(x_{e,i} \cdot S)
\end{split}
\end{equation}
where $W_s^T$, $\sigma$ are the weights of linear layer and sigmoid function. Note the second action, the pricing discount $p$, is output by a similar logic structure with 
$y_{rl}$; this part of logic is omitted in the figure, for simplicity. The left part of Figure \ref{fig:actor_critic} shows the actor structure introduced above.


\subsection{Critic}
\begin{figure}[t]
  \centering
  \includegraphics[width=1.0\linewidth]{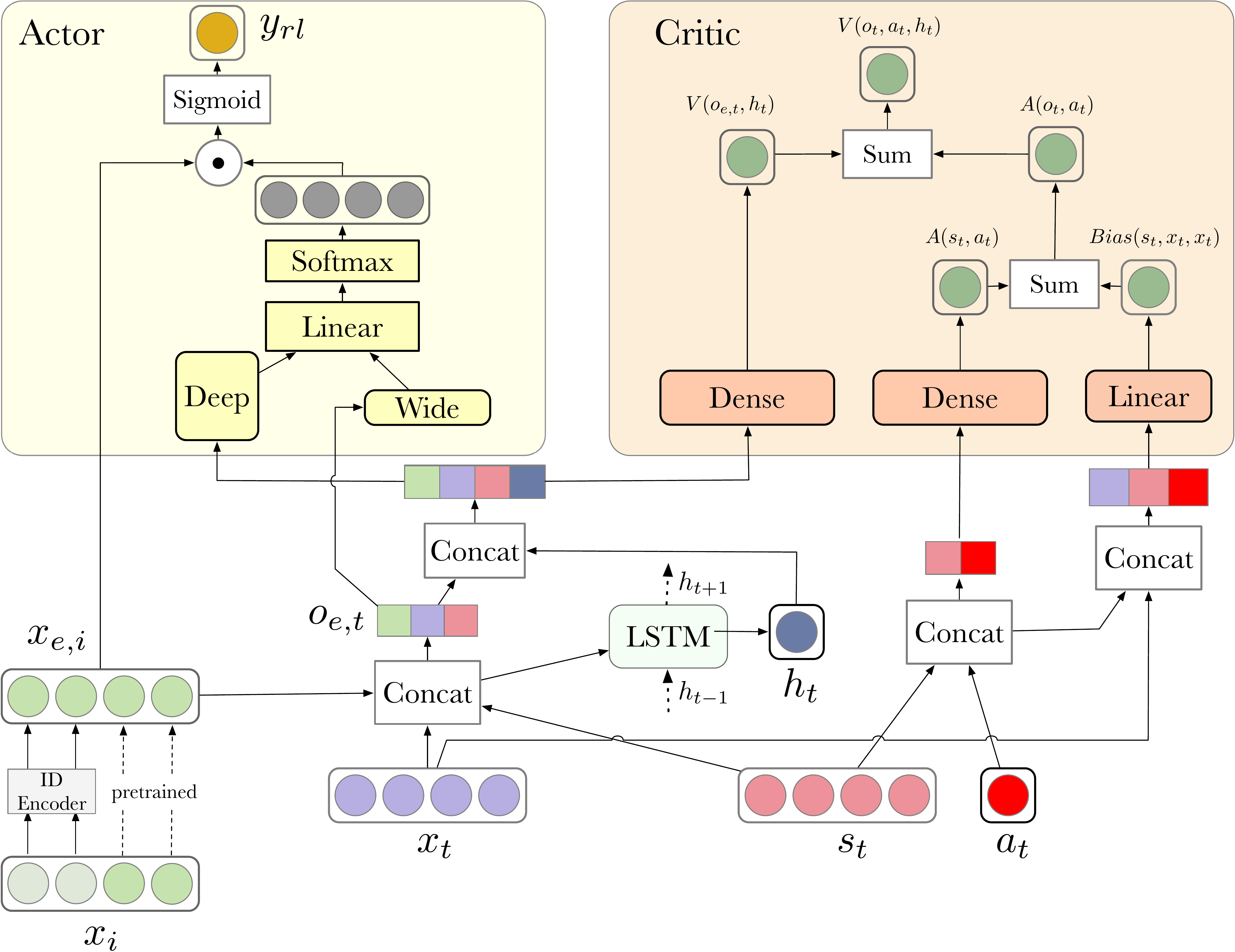}
  \caption{Our Actor-Critic Network Structure. Actor (left) and critic (right) share the same item encoder and LSTM component.} 
  \label{fig:actor_critic}
\end{figure}

The right part of Figure \ref{fig:actor_critic} illustrates our critic. It shares the same ID encoder and LSTM components with actor. A dueling structure \cite{Wang2016DuelingNA} is here employed to depict the action value function, by decomposing it into the state value $V(o)$ and the action advantage value $A(o, a)$. 
$V(o)$ is generate by a dense net given $o_{t}$ and $h_t$. For $A(o, a)$, we further decompose it as $A(o_t, a_t) = A(s_t, a_t) + Bias(s_t, x_t, a_t)$. $A(s_t, a_t)$ has the same meaning with what is defined in the vanilla dueling structure, which is calculated by another dense net; the second term, $Bias(s_t, x_t, a_t)$, representing the bias consideration caused by the trending-bias factor $x_t$, is calculated by a single linear layer. The final $Q$ value is obtained by simply adding all these terms: 

\begin{align}
    Q(o_t, a_t, h_t) &= V(o_{t},h_{t}) + A(o_t, a_t) \notag \\
    &= V(o_{e,t},h_{t}) + A(s_t, a_t) + Bias(s_t, x_t, a_t)
\end{align} 

\subsection{Learning}
RL aims to solve the MDP and provide a policy $\pi_{\theta}(a|s)$ which maximizes the expected discount return in Equation (\ref{eq:v}). Because our RL framework and online recommendation are in different time frequencies, an off-policy method is more suitable for us, such as DDPG introduced in Section \ref{sec:ddpg}. Same with the original version, here we also applies the double-Q network settings and a replay memory buffer. However, with the existence of the recurrent component, and also to reduce the computational burden for the industrial-scale data, here we consider a different way of data sampling and network update. In the vanilla DDPG method, one samples a batch of transitions (the tuple $(o_t, a_t, r_t, o_{t+1})$) and the gradient update is the batch average of each transition feedback. In our approach, we instead randomly sample a batch of items (or their episodes) from the buffer; updates are executed at the beginning of episode then proceed forward through time until the episode ends. In such manner, the hidden state in LSTM is always carried forward from its previous value, which alleviates its convergence speed. One can refer to \cite{SDMIA15-Hausknecht} for a similar update method in which it is called "Bootstrapped Sequential Updates". Note in our system, an episode is an item's transition sequence collected and sorted in the daily queue. Therefore, here we name our training algorithm as \textit{Itemwise Episodic Recurrent Deterministic Policy Gradient} (IE-RDPG) with the pseudo code summarized in Algorithm \ref{alg:ieddpg}.



\begin{algorithm}[t!]
\caption{IE-RDPG}
\label{alg:ieddpg}
\begin{algorithmic}[1]
\STATE {\textbf{Initialize} the parameters $w$ and  $\theta$ of critic $Q_w(o,a)$ and actor $\pi_{\theta}(a|o)$} 
\STATE {\textbf{Initialize} the target networks with $w^{\prime} \leftarrow w$ and $\theta^{\prime} \leftarrow \theta$}
\STATE {\textbf{Initialize} the replay memory buffer $R = \{ \}$} 
\STATE \textbf{REPEAT:} \\
\qquad \textbf{Transition Generation Stage:}
\STATE \qquad \textbf{for} item $ 1, M$ \textbf{do}
\STATE \qquad \qquad {Receive initial observation state $o_0$}
\STATE \qquad \qquad \textbf{for} $t = 1, T$ \textbf{do}
\STATE \qquad \qquad \qquad Perform $a_t$ using the current policy $\pi_{\theta}(a_t \vert o_t)$
\STATE \qquad \qquad \qquad Collect reward $r_t$ and the next observation $o_{t+1}$
\STATE \qquad \qquad \qquad {Record the transition $\mathcal{T}_t = (o_t, a_t, r_t, o_{t+1})$}
\STATE \qquad \qquad \textbf{end for}
\STATE \qquad \qquad Store the episode $\{\mathcal{T}_0, \mathcal{T}_1, \cdots, \mathcal{T}_T\}$ in $R$;
\STATE \qquad \textbf{end for}

\qquad \textbf{Parameter Updating Stage:}
\STATE \qquad \textbf{for} epoch $= 1, K$ \textbf{do}
\STATE \qquad \qquad {Sample a mini-batch of $N$ episodes from $R$}
\STATE \qquad \qquad \textbf{for} $t = 1, T$ \textbf{do}
\STATE \qquad \qquad \qquad {Set $y_i = r_i + \gamma Q_{w^{\prime}}(o_{i+1}, \pi_{\theta^{\prime}}(a_{i+1} \vert o_{i+1}))$}
\STATE \qquad \qquad \qquad {Update critic by minimizing the loss: \\
\qquad \qquad \qquad $L = \frac{1}{N} \sum_i (y_i - Q_w(o_i, a_i))^2$}
\STATE \qquad \qquad \qquad Update actor using the sampled policy gradient: \\
\qquad \qquad \qquad $\nabla_{\theta} J =  \frac{1}{N} \sum_i \nabla_a Q_w(o, a) \nabla_{\theta} \pi_{\theta}(a \vert o)$
\STATE \qquad \qquad \qquad {Update the target networks:}
\begin{align}
&w^{\prime} \leftarrow \tau w + (1 - \tau) w^{\prime} \notag \\
&\theta^{\prime} \leftarrow \tau \theta + (1 - \tau) \theta^{\prime} \notag
\end{align}
\vspace{-5mm}
\STATE \qquad \qquad \textbf{end for}
\STATE \qquad \textbf{end for}
\end{algorithmic}
\end{algorithm}

There are two phases within a training session of IE-RDPG. First, we let the agent to interact with the platform with respect to the current policy, collect the data until $T$ days. For each item, transition is stored in chronological order to form the episode. After that, the parameter updating stage starts with randomly sampling a batch of $N$ episodes at one time. The parameter update is first conducted parallelly at the beginning timestep, then such update continues to proceed forward through time until all episodes end. The parameter updating stage could have multiple epochs. This two-phase session repeats in the real-world timeline. IE-RDPG provides an asynchronous and distributed architecture which is able to solve the problem with enormous items in the one of the largest E-commerce platform. 


\subsection{Dual-rank module} 
\label{subsec: dual_rank}
Pointwise Learning to Rank (LTR) is widely used in industrial recommendation ranking. Upon each user request, the item feature is aligned with the user characteristic and an user-affinity score is calculated by a supervised model like click-through rate (CTR) model. This ctr score, 
$y_{ctr}$, is  employed as the ranking metric to yield a ranked item list. In our study, RL-LTV provides another item-affinity score, $y_{rl}$, 
indicating its preference based on the item LTV. By forming a dual-rank module and considering these two scores simultaneously, the online ranking could consider not only the immediate click-through reward but also the long-term rewards. As a result, some fresh items without enough historical interactions but with high potential of LTV, would have small $y_{ctr}$ but large $y_{rl}$, and is expect to have more ranking priority.


We obtain the new ranking score by the following equation:
\begin{equation}
\label{eq:dual_rank}
    y = (1-\alpha) \cdot y_{ctr}+ \alpha \cdot y_{rl}
\end{equation}
where $\alpha \in (0, 1)$ is the mixing parameter. Since $Q$ output by the critic predicts the future LTV, it could be used to determine the degree of $\alpha$, therefore also determine how the CTR and LTV rewards tradeoff. In the experiments, $\alpha$ is calculated as
\begin{equation}
\label{eq:alpha}
    \alpha = e^{\frac{Q-Q_{min}}{Q_{max} - Q_{min}} \ln(1+\alpha_{\text{max}}-\alpha_{\text{min}})} - 1 + \alpha_{\text{min}}
\end{equation}
where $\alpha_{\text{min}},\alpha_{\text{max}}$ are hyper-parameters highlighting the lower and upper bounds of $\alpha$; and item with higher $Q$ would always have a higher $\alpha$.

\section{EXPERIMENTS}
\label{sec:experiment}

To evaluate the proposed model, we apply our approach on Taobao, a world-leading E-commerce platform. A series of experiments are designed in order to address the following questions: (1) whether the proposed approach performs better than baseline, or some other competitors which is not RL-based; (2) verify if inclusion of $x$, or the LSTM component contributes to the performance; and (3) if some representative examples could be found to display how the framework improve LTV, from the business point of view. 
It is important to emphasize that,  most of public dataset in the recommendation domain do not have the characteristic of item dynamics depicted in Section \ref{sec:item_growth}, therefore we can not simply directly comparing our performance with most state-of-the-art baselines. On the other hand, the live environment contains billions of users and items, therefore it is also not a trivial task to launch an arbitrary baseline on the real system. As a remedy, the performance of RL-LTV is comparing with the live ranking algorithm (refer to \cite{10.1145/3219819.3219823} for part of details) which is a reasonable baseline for such a complicated industrial system. We also conduct offline analysis on the LTV recognition including comparison of several baselines in Section \ref{subsec:offline_test}. Ablation and sensitivity analysis are also provided to further illustrate the reasonability of our framework.

\subsection{Experimental Settings}


\textbf{Training process}: The online ranking system is in real-time, while our approach has a daily update frequency. In our practice, user logs are kept collecting, then data is aggregated by item and by day, and the MDP transitions are recorded. According to Algorithm \ref{alg:ieddpg}, for each item we wait for $T$ new transitions to appear; a new episode is then formed and put into the buffer. Similarly, it is the episodes from each buffer sampling, and gradient update is performed accordingly. 

\textbf{Parameters setting}: In the training of our RL-LTV, we use Adam with learning rate 0.0001 as the optimizer to train actor and critic. The hyper-parameters $\gamma$, $\tau$, $\alpha_{min}$ and $\alpha_{max}$ are 0.5, 0.001, 0 and 0.2, respectively. The dimension of hidden states in the LSTM component is 4. The item inherent feature output by the item encoder has the dimension of 8. From the applicable point of view, the time step is chosen as a day. We have relatively small buffer (200) and batch size (50), because our replay memory buffer is operated in terms of episodes, not transitions. 



\subsection{LTV Recognition}
\label{subsec:offline_test}

Because recommendation is directly correlated with customer experience and platform revenue, scale and depth of the online experiment are limited at the current stage. Nevertheless,, some offline analysis can be conducted from the perspective of LTV recognition.

\textbf{Metrics:} Two aspects of performances could be evaluated offline: (1) Since the critic provides a predicted version of the actual cumulative gain, $Q$, the regression accuracy between $Q$ and the actual $J$, $J_{\text{actual}}$, can be calculated and evaluated; (2) The actor generates $y_{rl}$ which indicates its ranking preference \textit{w.r.t.} LTV; a ranking metric can be evaluated by sorting $y_{rl}$ in a descending order, and comparing its sequential similarity with the ground truth (item sorted by $J_{\text{actual}}$).

To calculate $J_{\text{actual}}$, data of 5 consecutive days are collected; while data in longer horizons can be ignored since their weights in $J$ decay quickly with $\gamma = 0.5$. For (1), the prediction error of this $J_{\text{actual}}$ is studied in the manner of root mean square error (RMSE) and mean absolute error (MAE)  .

We employ the normalized discounted cumulative gain (NDCG) as a measure of the sequential similarity stated in (2). To mimic the live environment and reduce the computation cost, the online retrieval method is also employed for this offline evaluation, which generates a series of items upon each user query. The top-$K$ items from the retrieved list are used to calculate NDCG@K with $K = 10, 20, 50$, and $J_{\text{actual}}$ is used as the relativity score. 


\textbf{Baselines:} The following experimental versions are employed to compare with our RL-LTV: 

\begin{itemize}
    \item \textit{Vanilla-CTR}: The online, pointwise, single-period, vanilla CTR model provides a natural baseline running everyday on Taobao, with a state-of-the-art CTR performance by so far but without any LTV consideration. 
    \item \textit{Empirical}: To mimic the human decision making process, scores are manually assigned based on business experience, \emph{e.g.} collect percentiles of $J_{\text{actual}}$ according to category, seller and brand respectively, calculate their weighted averages, then scale to (0,1).
    \item \textit{LSTM}: Use a supervised LSTM model to regress and predict $J_{\text{actual}}$. Score is then assigned based on prediction. It shares the same structure as the LSTM component in RL-LTV.
\end{itemize}

Note we only compare the regression accuracy of LSTM with RL-LTV, while Vanilla-CTR or Empirical do not have an explicit LTV prediction. On the other hand, we can evaluated the ranking NDCG of LTV for all these baselines. For Vanilla-CTR, the CTR score is used to rank its preference for LTV ranking, which is actually the exact case on the live platform. 

\begin{table}
  \caption{Offline metrics on item LTV. Note Baseline and Empirical do not have RMSE or MAE result because they do not have $Q$ estimates.}
  \label{tab:offline_accuracy}
  \scalebox{0.85}{
    \begin{tabular}{cccccc}
        \toprule
        Model & RMSE & MAE & NDCG@10 & NDCG@20 & NDCG@50\\
        \midrule
        Vanilla-CTR & $--$ &  $--$  & $0.746$ & $0.662$ & $0.582$ \\ 
        \hline
        Empirical &  $--$  &  $--$  & $0.765$ & $0.678$ & $0.581$ \\ 
        \hline
        LSTM & $7.900$ & $4.761$ & $0.772$ & $0.684$ & $0.600$ \\ 
        \hline
        RL-LTV & $7.873$ & $4.067$ & $0.782$ & $0.705$ & $0.625$ \\
      \bottomrule
    \end{tabular}
  }
 \vspace{-2mm}
\end{table}

\textbf{Result:} We summarize these results in Table \ref{tab:offline_accuracy}. RL-LTV is expected to have superior LTV regression accuracy, as a result of its interactive learning behavior and long time-horizon optimization, which is verified by the result. Compared to LSTM, RL-LTV reduces the prediction error of LTV for both MAE and RMSE. For NDCG@K, one can find that Empirical, LSTM and RL-LTV all have better NDCGs than Vanilla-CTR, which only looks at the instant CTR metric. Still, RL-LTV is the best among these experiments.

\subsection{Live Experiments}
\label{subsec:online_test}

The live experiment is conducted on Taobao. The A/B test starts on January 26th, 2021 and lasts for a week. Policy of RL-LTV is first warmed up with the trained version in Section \ref{subsec:offline_test}, then updates online in a daily basis. Because of the real world limitation, only 
$y_{rl}$ in the action takes effective; $p$ is instead determined and controlled by other group, therefore becomes a read-only parameter for us.

\textbf{Cold-start performance:} We first investigate the performance of cold-start items. To define cold-start items, fresh products with time on the market less then a month when the experiment starts are tagged, and then randomly divided into different groups. Each group of cold-start items is limited to be recommended by only one experiment, such that the item-based metrics (IPV and GMV) of different experiments can be reasonably calculated and fairly compared. LTV in forms of PV, IPV and GMV are collected after the online test ends. For RL-LTV, relative LTV differences to Vanilla-CTR are calculated and shown in Table \ref{tab:perf_cold}. One can see that RL-LTV improves LTVs of IPV and GMV significantly, with almost the same PV investment. 

\begin{table}
  \caption{Percentage Difference of LTV relative to Vanilla-CTR for cold-start items.}
  \label{tab:perf_cold}
  \vspace{-1mm}
\scalebox{0.85}{
  \begin{tabular}{cccc}
    \toprule
    Model &PV &IPV &GMV \\
    \midrule
    \hline
    RL-LTV (w/o R) &$1.01\%$ &$6.25\%$ &$12.31\%$ \\
    \hline
    RL-LTV &$-1.19\%$ &$8.67\%$ & $18.03\%$ \\
    \bottomrule
  \end{tabular}
}
\vspace{-2mm}
\end{table}

\begin{table}[t]
\center
\caption{Comparison of Offline Performance for Component Ablation Analysis} 
\label{tab:ablation}
\vspace{-1mm}
\scalebox{0.85}{
\renewcommand{\arraystretch}{1}
\begin{tabular}{c|c|c|c|c|c}
\toprule
\multicolumn{1}{c|}{Model}&
\multicolumn{1}{c|}{RMSE} & \multicolumn{1}{c|}{MAE} &  \multicolumn{1}{c|}{NDCG@10} &\multicolumn{1}{c|}{NDCG@20} & \multicolumn{1}{c}{NDCG@50}\\
\hline
	RL-LTV (w/o $x_s$)& $7.947$& $4.135$ &$0.771$ & $0.686$ & $0.613$  \\
\hline
	RL-LTV (w/o $x_t$) & $8.146$ &$4.351$ &$0.763$ & $0.680$ & $0.611$ \\
\hline
	RL-LTV (w/o R) &$7.931$ &$4.133$ &$0.754$&$0.672$ & $0.592$ \\
\hline
    RL-LTV &$7.873$ &$4.067$ &$0.782$ & $0.705$  &$0.625$\\
	\bottomrule
	\end{tabular}
	\label{tab:dod}
}
\vspace{-3.5mm}
\end{table}

\textbf{Global performance:} The ultimate purpose of cold-start recommendation is to improve the entire RS performance. Therefore, we need to inspect metrics of all items, not only those of cold-start items. Although the short-term performance of RS might be harmed due to the investment on cold-start items, it is expected its long-term performance could be reimbursed with the growth of LTVs. By the end of experiment, such reimbursement effect is observed. In more details, RL-LTV has $-0.55\%$ PV and $-0.60\%$ IPV compared with Vanilla-CTR, indicating there is no severe degradation for the global performance.

\begin{figure}[t!]
  \centering
  \includegraphics[width=7.5cm]{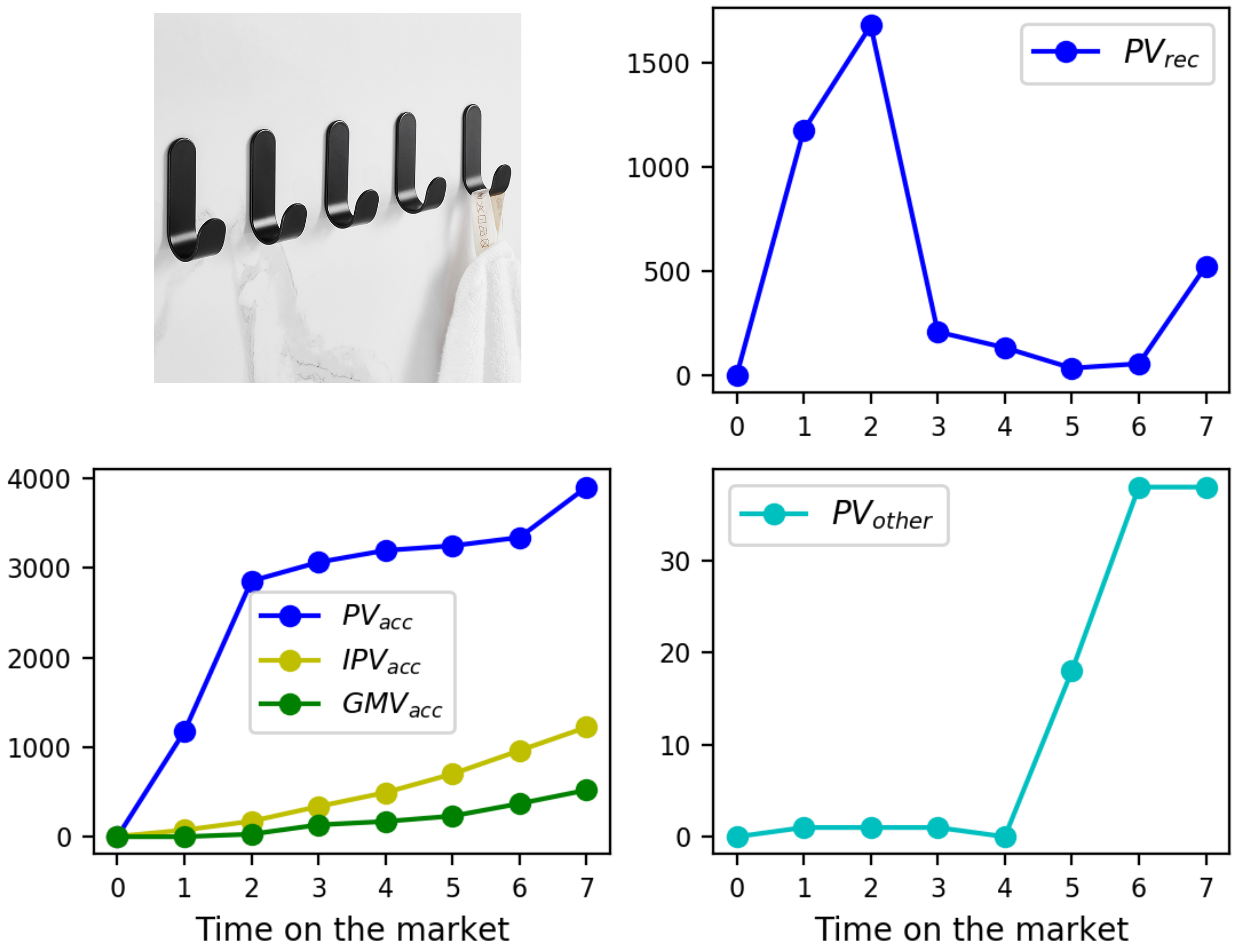}
  \caption{Time trajectory of a typical cold-start recommended item (A hook handle as a fresh item of platform). Upper Right: The $PV_{\text{rec}}$ invested by RL-LTV. Lower Right: $PV_{\text{other}}$ (mainly from the search channel) inspired after LTV investment. Lower Left: The accumulated PV, IPV and GMV indicating this item rapidly becomes a star-item.}
  \label{fig:item_age_metric}
\end{figure}

\textbf{Typical case:} In order to have a deeper insight that how RL helps recognize a specific high potential item and improve its LTV, we also look into some specific cases. Figure \ref{fig:item_age_metric} shows such a case study with an item (a hook handle) first introduced on the platform on January 26th. As it cold starts, there is few people who comes to view, click and buy it. However, RL-LTV recognizes it as high LTV potential, therefore makes substantial PV investment at the early stage (the $PV_\text{rec}$ curve on the upper right). This investment is successful since it triggers more people's interests and behaviors (the $PV_\text{other}$ (PV from the other channels) curve on the lower right), with PV, IPV and GMV all grow rapidly (the accumulated curves on the lower left). After this item turns into a star product, RL-LTV turns to invest other cold-start items. As a result, $PV_\text{rec}$ falls down after the 3rd day, while the other metrics are already in a healthy closed-loop feedback.

\subsection{Ablation Studies}
To illustrate the effectiveness of several important components in RL-LTV, we here perform some ablation analysis on the follow attempts:
\begin{itemize}
    \item RL-LTV(w/o $x_s$): our approach but without the inclusion of the item inherent feature, $x_s$.
    \item RL-LTV(w/o $x_t$): our approach but without the inclusion of the trending-bias factor, $x_t$.
    \item RL-LTV(w/o R): our approach but without the recurrent LSTM component.
\end{itemize}
Similar to Section \ref{subsec:offline_test}, we perform the offline analysis for these experiments, and their results aside with RL-LTV are given in Table \ref{tab:ablation}. Not surprisingly, RL-LTV still has the best performance, suggesting that $x_s$, $x_t$ and the recurrent cell are all crucial. Due to limited online resource, only RL-LTV(w/o R) is conducted with an live experiment, with result also shown in Table \ref{tab:perf_cold}. One can see that RL-LTV(w/o R) also has positive online impacts compared with Vanilla-CTR, but not as good as RL-LTV.

All these results emphasize the validity of our original definition of POC-MDP. Comparison with RL-LTV, the degraded performance of Vanilla-CTR (in Subsection \ref{subsec:offline_test} and \ref{subsec:online_test}) verifies the entire MDP and RL framework; the degraded performances of RL-LTV(w/o $x_s$) and RL-LTV(w/o $x_t$) indicates the necessity of uncontrollable state (\emph{i.e.}, $x_s$ and $x_t$); the degraded performance of RL-LTV(w/o R) suggests the unobservable state also helps the modeling.

\subsection{Hyperparameter Sensitivity Analysis}
\begin{figure}[t!]
  \centering
    \includegraphics[width=7.5cm]{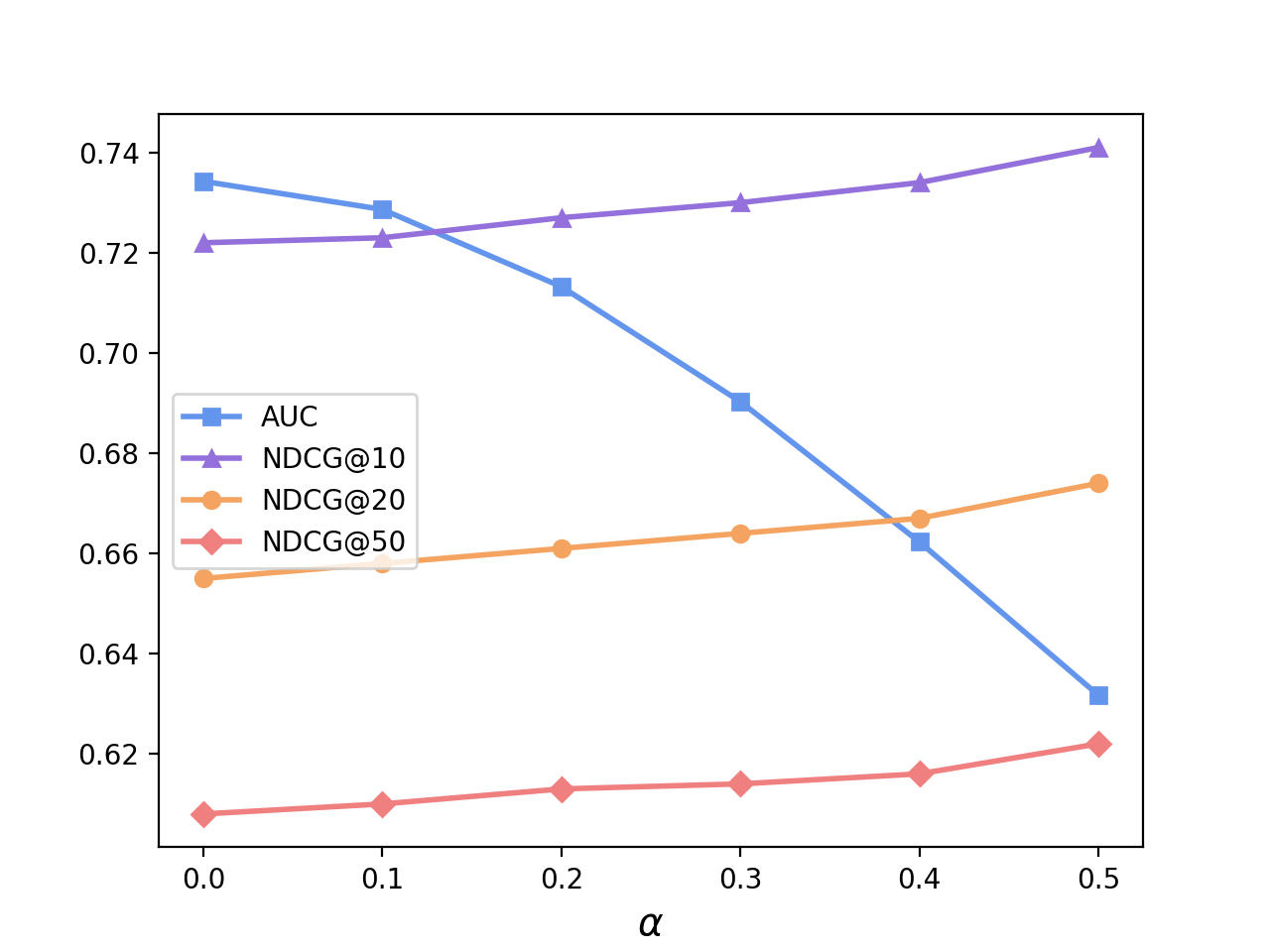} 
  \caption{AUC, NDCG@10, NDCG@20 and NDCG@50 curves with different $\alpha$. When more proportion of $y_{rl}$ is adopted against $y_{ctr}$, AUC decreases but NDCG of LTV increases, as expected.}
  \label{fig:alpha_metric}
\vspace{-4mm}
\end{figure}

Here we also perform a sensitivity analysis for an important parameter, the incorporation weight $\alpha$ in Equation (\ref{eq:dual_rank}). Different choices of $\alpha$ generates different final score $y$, which similarly, can also have an NDCG evaluation introduced in Section \ref{subsec:offline_test}. Furthermore, as an indication of CTR prediction with the true CTR label, AUC of $y$ can also be evaluated with respect to different $\alpha$. As $\alpha$ steadily changes from 0 to 1, $y$ transforms from a ranking metric of CTR to a ranking metric of item LTV. By checking the curves of AUC (of CTR) and NDCG (of LTV) w.r.t. $\alpha$, one could have a snapshot about the trade-off between the instant reward and the long-term reward. \footnote{Note $\alpha$ should vary with different items according to Equation (\ref{eq:alpha}). However, here we just perform a simplified offline analysis by fixing all items with the same $\alpha$.} The result is demonstrated in Figure \ref{fig:alpha_metric}. Not surprisingly, the AUC decreases while NDCG increases as $\alpha$ increases. Based on the shape of curves, we expect the mean of $\alpha$ to be around 0.1 where the AUC curve starts to decrease faster. Correspondingly, $\alpha_{\text{min}}$ and $\alpha_{\text{max}}$ in Equation (\ref{eq:alpha}) are set to 0 and 0.2, respectively.

\section{CONCLUSION}
\label{sec:conclusion}
In this paper, we propose a novel RL-LTV framework which solves the cold-start recommendation problem by considering the longer-term rewards of items, the LTVs. A special form of MDP, POC-MDP, is employed to model the item life dynamics. Generalized item-level observation and action spaces, as well as parameter-shared networks, help the knowledge transfer from mature, historical items to fresh, cold-start items. An off-policy, actor-critic RL agent with an LSTM component is employed to interactively learn and change the online system with a policy optimizing both instant reward and long-term LTV. It is the first time to incorporate such policies into the online recommendation system, to address the item cold-start issue. We develop a training framework called IE-RDPG to complete the large scale, itemwise episodic training task. By rigorous experiments, it shows that our algorithm performs much better on the long-term rewards of cold-start items, comparing with the online state-of-the-art baseline on Taobao. There is no evident degradation of the entire online performance during the experiment. For future work, it would be interesting to use RL to study the entire life periods of products, therefore produce policies not only for cold-start but also lifelong recommendation. Another possible improvement is to make the model more explainable to explicitly exhibit the lifetime stage transitions of products.

\balance
\bibliographystyle{ACM-Reference-Format}
\bibliography{main}

\end{document}